\documentclass[twoside,fleqn]{article}
\usepackage[dvips]{graphics}
\usepackage{espcrc2}
\usepackage{epsfig}
\usepackage{latexsym}

\newcommand{\bq}{\begin{equation}}
\newcommand{\eq}{\end{equation}}
\newcommand{\bqa}{\begin{eqnarray}}
\newcommand{\eqa}{\end{eqnarray}}
\newcommand{\bqas}{\begin{eqnarray*}}
\newcommand{\eqas}{\end{eqnarray*}}
\newcommand{\bdm}{\begin{displaymath}}
\newcommand{\edm}{\end{displaymath}}

\newcommand{\plaq}{\mbox{\raisebox{-2.4mm}
{\epsfig{file=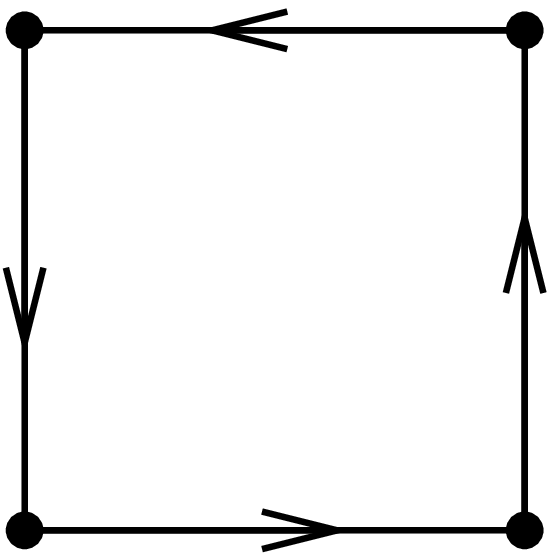,height=6mm
}}~}}
\newcommand{\loopa}{\mbox{\raisebox{-2.mm}
{\epsfig{file=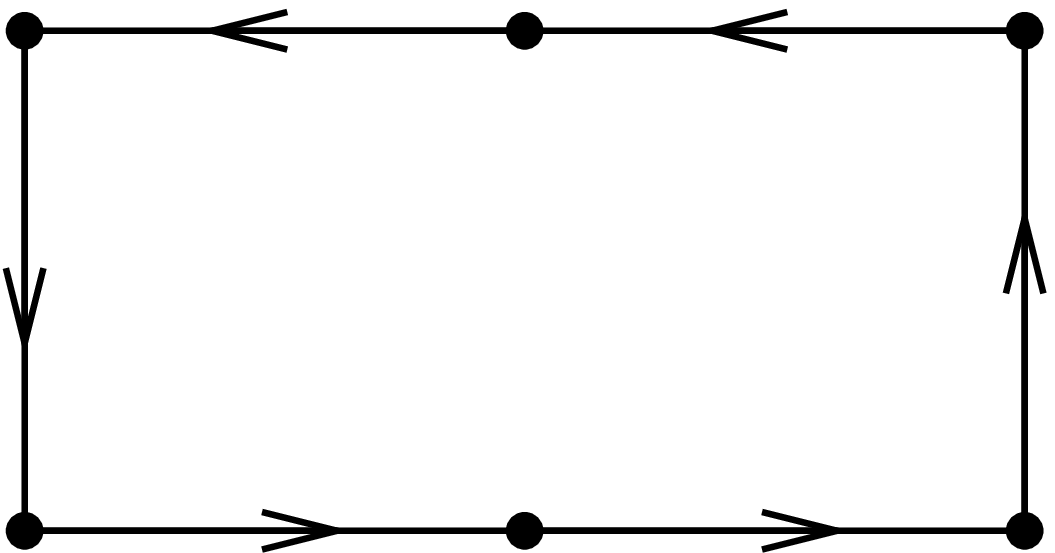,height=6mm
}}~}}
\newcommand{\loopb}{\mbox{\raisebox{-4mm}
{\epsfig{file=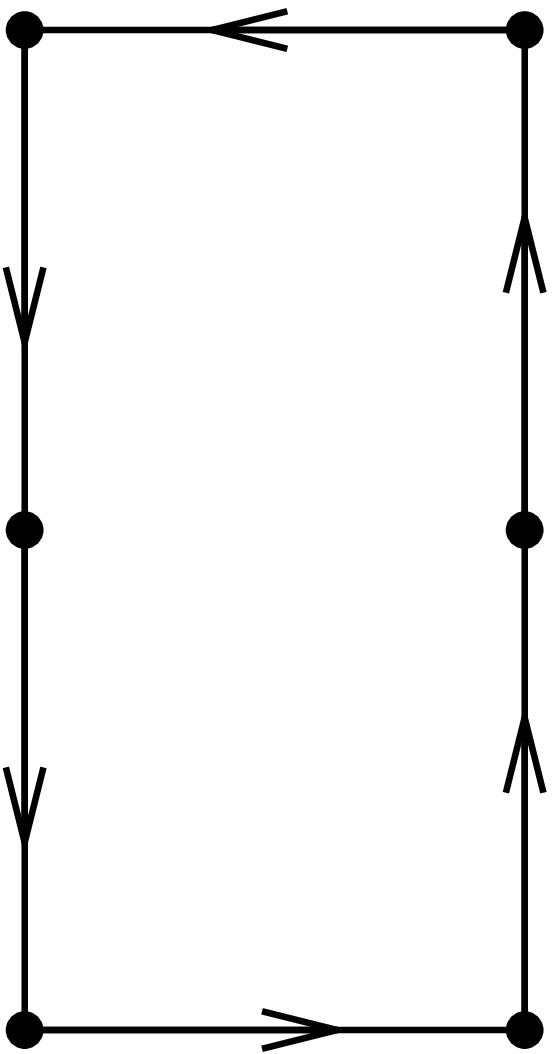,height=12mm
}}~}}
\newcommand{\fatlink}{\mbox{\raisebox{-0.15mm}
{\epsfig{file=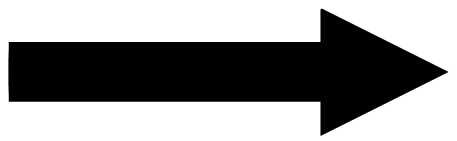,height=1.1mm
}}~}}
\newcommand{\fatlinka}{\mbox{\raisebox{-0.15mm}
{\epsfig{file=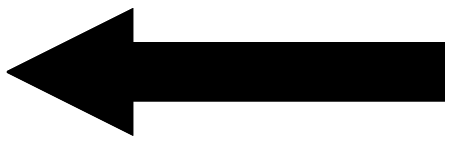,height=1.1mm
}}~}}
\newcommand{\alink}{\mbox{
\begin{picture}(2.5,.2)
\linethickness{1mm}
\multiput(0,0.1)(2,0){2}{\circle*{0.1}}
\multiput(1.01,0.1)(0,0){1}{\circle{0.2}}
\put(1,0.){\fatlink}
\put(0,0.){\fatlinka}
\put(1,-.5){\scriptsize \( i \) }
\put(-.5,-.5){\scriptsize \( j \) }
\put(1.9,-.5){\scriptsize \( j \) }
\end{picture}}}
\newcommand{\blinkba}{\mbox{
\begin{picture}(2.5,2.5)
\thicklines
\multiput(1,-1)(0,-1){2}{\circle*{0.1}}
\multiput(1,0)(0,0){1}{\circle{0.2}}
\multiput(2,0.0)(0,1){3}{\circle*{0.1}}
\multiput(0,-2)(0,0){1}{\circle*{0.1}}
\put(1,-2.0){\vector(-1,0){1}}
\put(1,-1.0){\vector(0,-1){1}}
\put(1,0.0){\vector(0,-1){1}}
\put(1,0.0){\vector(1,0){1}}
\put(2,0.0){\vector(0,1){1}}
\put(2,1.0){\vector(0,1){1}}
\put(0.5,-0.1){\scriptsize \( i \) }
\put(1.5,2.0){\scriptsize \( j \) }
\put(-0.4,-2.1){\scriptsize \( j \) }
\end{picture}}}
\newcommand{\blinkbb}{\mbox{
\begin{picture}(2.5,2.5)
\thicklines
\multiput(0,0.0)(0,-1){3}{\circle*{0.1}}
\multiput(1,1.0)(0,1){2}{\circle*{0.1}}
\multiput(1,0.0)(0,){1}{\circle{0.2}}
\multiput(2,2)(0,1){1}{\circle*{0.1}}
\put(0,-1.0){\vector(0,-1){1}}
\put(0,0){\vector(0,-1){1}}
\put(1,0){\vector(-1,0){1}}
\put(1,0){\vector(0,1){1}}
\put(1,1){\vector(0,1){1}}
\put(1,2){\vector(1,0){1}}
\put(1.25,-0.1){\scriptsize \( i \) }
\put(2.2,2.0){\scriptsize \( j \) }
\put(-0.4,-2.1){\scriptsize \( j \) }
\end{picture}}}
\newcommand{\blinkbc}{\mbox{
\begin{picture}(2.5,2.5)
\thicklines
\multiput(1,1)(0,1){2}{\circle*{0.1}}
\multiput(1,0)(0,0){1}{\circle{0.2}}
\multiput(2,0)(0,-1){3}{\circle*{0.1}}
\multiput(0,2)(0,1){1}{\circle*{0.1}}
\put(1,2){\vector(-1,0){1}}
\put(1,1){\vector(0,1){1}}
\put(1,0){\vector(0,1){1}}
\put(1,0){\vector(1,0){1}}
\put(2,0){\vector(0,-1){1}}
\put(2,-1){\vector(0,-1){1}}
\put(0.5,-0.1){\scriptsize \( i \) }
\put(1.5,-2.1){\scriptsize \( j \) }
\put(-0.5,2.0){\scriptsize \( j \) }
\end{picture}}}
\newcommand{\blinkbd}{\mbox{
\begin{picture}(2.5,2.5)
\thicklines
\multiput(0,0)(0,1){3}{\circle*{0.1}}
\multiput(1,-1)(0,-1){2}{\circle*{0.1}}
\multiput(1,0)(0,-1){1}{\circle{0.2}}
\multiput(2,-2)(0,1){1}{\circle*{0.1}}
\put(0,1){\vector(0,1){1}}
\put(0,0){\vector(0,1){1}}
\put(1,0){\vector(-1,0){1}}
\put(1,0){\vector(0,-1){1}}
\put(1,-1){\vector(0,-1){1}}
\put(1,-2){\vector(1,0){1}}
\put(1.25,-0.1){\scriptsize \( i \) }
\put(2.1,-2.1){\scriptsize \( j \) }
\put(-0.5,2.0){\scriptsize \( j \) }
\end{picture}}}

\newfont{\mega}{cmr17 scaled 2000}
\newfont{\Mega}{cmr17 scaled 3000}
\newfont{\MEGA}{cmr17 scaled 4000}
\newfont{\giga}{cmr17 scaled 5000}
\newfont{\Giga}{cmr17 scaled 6000}
\newfont{\GIGA}{cmr17 scaled 8000}

\title{ The three flavour chiral phase transition with an improved quark
    and gluon action in lattice QCD \thanks{ This work was partly supported by
    the Deutsche Forschungsgemeinschaft under grant Ka 1198/3-1 and the EU TMR
    network grant ERBFMRX-CT97-0122.}}

\author{ A. Peikert with F. Karsch, E. Laermann and B. Sturm\\[3mm]Fakult\"at
    f\"ur Physik, Universit\"at Bielefeld, 33501 Bielefeld, Germany}

\begin{document}

\begin{abstract}
The finite-temperature chiral phase transition is investigated for three
flavours of staggered quarks on a lattice of temporal extent $N_\tau=4$.
In the simulation we use an improved fermion action which reduces rotational
symmetry breaking of the quark propagator ( p4-action ), include fat-links to
improve the flavour symmetry and use the tree level improved (1,2) gluon
action. We study the nature of the phase transition for quark masses of
$ma=0.025$, $ma=0.05$ and $ma=0.1$ on lattices with spatial sizes of $8^3$ and
$16^3$.
\end{abstract}

\maketitle

\section{INTRODUCTION}
QCD predicts a finite-temperature chiral phase transition from a hadronic to a
plasma phase. The nature of this phase transition depends on the number of
flavours and the quark masses.
Results from linear sigma models suggest\cite{wilczek92}
a second order transition for two and a first order transition for
three degenerate quarks in the limit of zero quark mass. The second order
 \vspace*{4mm} line
\epsfig{file=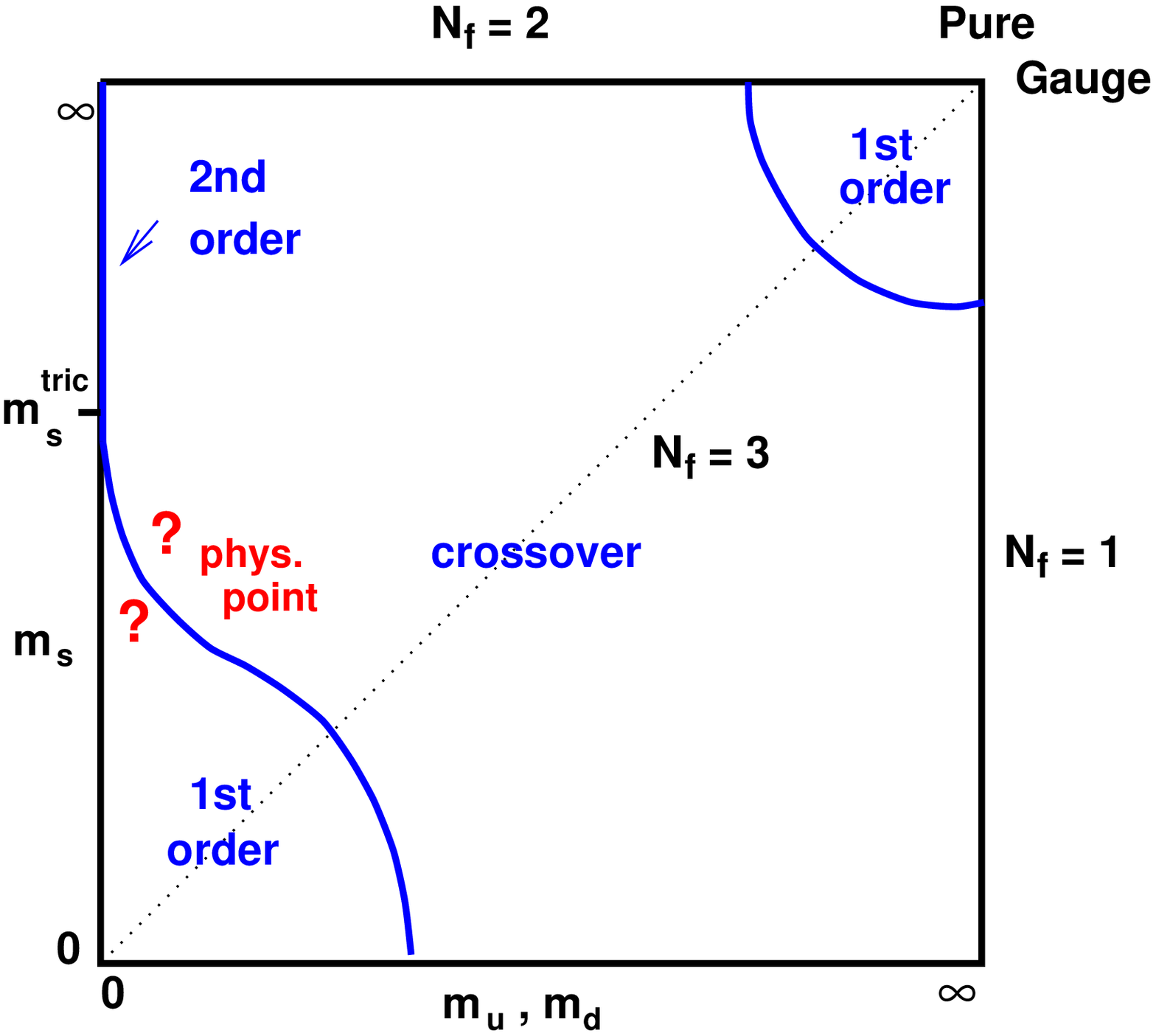,height=63mm,width=73mm}
{\vspace*{0mm}{~\newline}}
{ Figure 1. The nature of the finite-temperature QCD phase transition as
  a function of $m_{u,d}a$ and $m_sa$.}\\
\epsfig{file=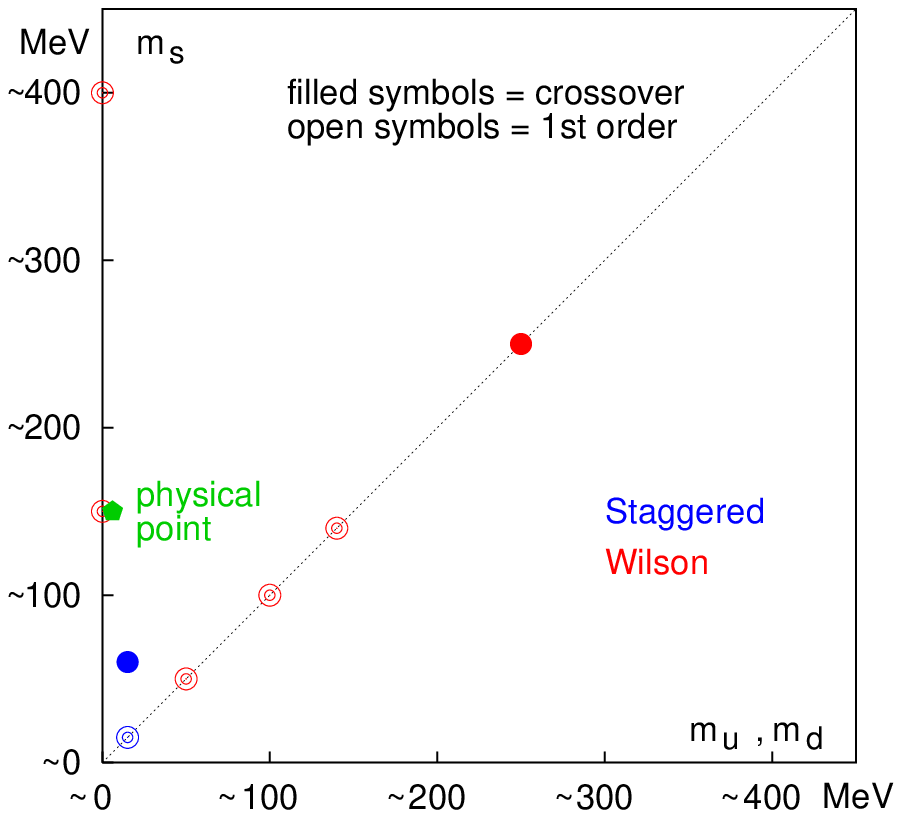,height=66.0mm,width=99.mm}
{\vspace*{-3mm}{~\newline}}
{ Figure 2. Simulation results for the order of the QCD phase transition
  for $N_f=2+1$ staggered\cite{brown90} and Wilson\cite{iwasaki96} fermions
  for $N_{\tau}=4$.}\\[2mm]
 in figure 1 runs into a tricritical point and then
 continues according to  $m_u,m_d \propto \left( m_s^{tric} - m_s \right)^{5 \over 2}$
separating the first order from the crossover region.
In figure 2 these predictions are compared with results from lattice simulations
using staggered\cite{brown90} and Wilson\cite{iwasaki96} fermions.
These data lead to different results for the order of the
phase transition at the physical point of two light up and down quarks and one
heavier strange quark. Both simulations may however suffer from large deviations from the
continuum limit which can be reduced by using an improved action.

\section{THE ACTION}
We used the $1 \times 2$ action for the gauge fields
\bqas
S_G=\beta \sum_{x, \nu > \mu} {5\over 3}\Bigg(
1-\frac{1}{N}{\rm Re Tr}~\plaq_{\mu\nu}\!\!\!(x)\Bigg)~-~~~~~~ \nonumber\\
\frac{1}{6}\Bigg(1-\frac{1}{2N}{\rm Re Tr}
\Bigg(\loopa_{\mu\nu}\!\!(x)+\loopb_{\mu\nu}\!\!(x)\Bigg)\Bigg), \nonumber\\[-3mm]
\eqas
which leads to an improved finite cut-off behaviour for quantities like
latent heat, surface tension and  pressure in pure SU(3) gauge
simulations\cite{beinlich96}.
For the fermion fields we used the p4-action which is constructed from
1-link and L-shaped 3-link paths in the derivative under the constraint that
the free quark propagator should be rotationally invariant up to order $p^4$.
The appropriate coefficients have been calculated at tree-level and to
one-loop-order\cite{heller98}. The off-diagonal part of the fermion matrix is
then given by
\bqas
M[U]_{ij}=\eta_{i} \Bigg\{ \left({3 \over 8} +
  g^2~{N^2-1\over 2N} ~0.0165 \right) ~A[U]_{ij} ~~\\
+ \left({1 \over 48} - g^2~{N^2-1\over 2N}~{1\over 6}~0.0165 \right)
~{1\over2}~B[U]_{ij} \Bigg\}
\eqas
\vspace*{-2mm}
\bqas
A[U]_{ij}\!&\!\!\!\!=&\!\!\!\alink\\[3.mm]
B[U]_{ij}\!&\!\!\!\!=&\!\!\!\blinkbd + \!\! \blinkbc ~+ \blinkba + ~\blinkbb.\\[2mm]
\eqas
The improvement of rotational symmetry leads to a fast approach to the
continuum ideal gas\footnotetext[1]{The Naik action also has a propagator
  invariant up to order $p^4$.} limit
as can be seen for the free energy in figure 3.
Both p4 and Naik$^1$-action  are much
closer to the ideal gas value already at small $N_\tau$ than the standard staggered action.
In figure 4 the effect of improving the p4-action to one-loop order is shown.\\
In addition fat-links\cite{blum97} are used in the one-link derivative $A[U]_{ij}$
which leads to an improved flavour symmetry indicated by a reduced pion
splitting. This effect has been seen for the tree-level p4-action in a
quenched \vspace*{2mm} simulation\cite{peikert98}.
\epsfig{file=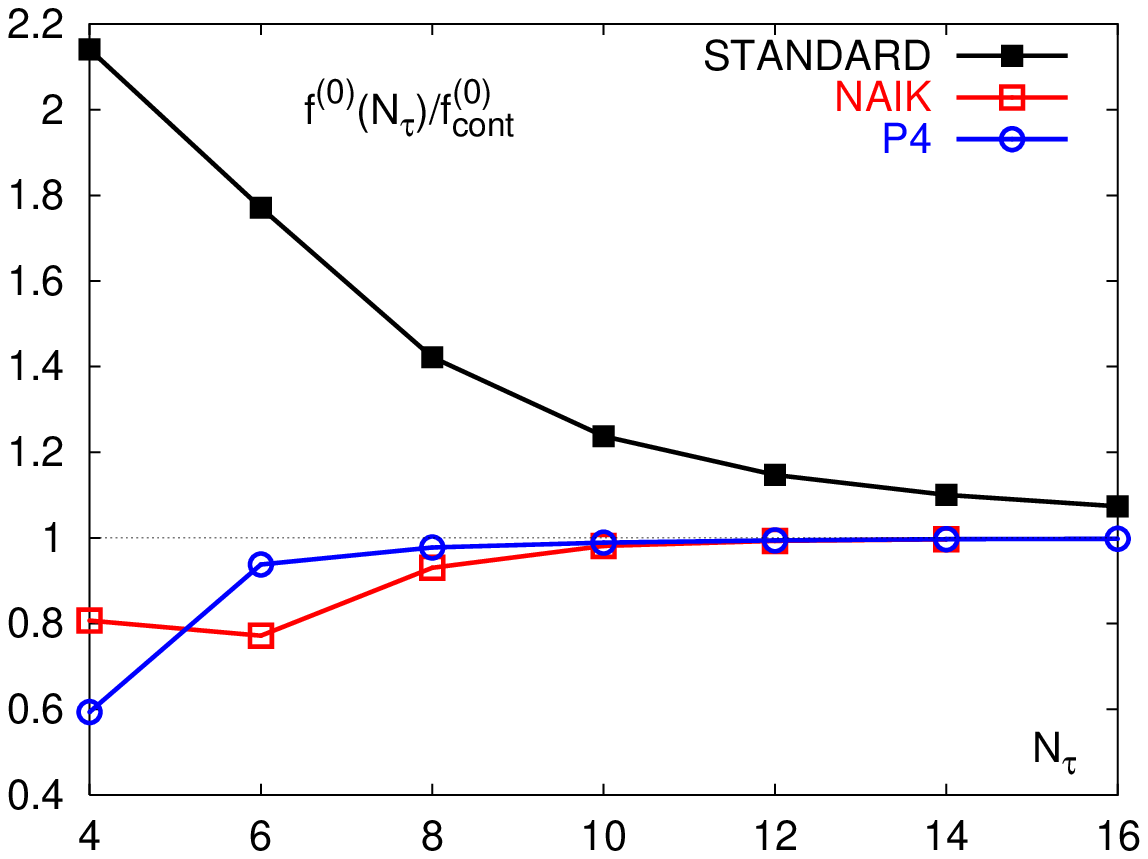,width=75mm}
{Figure 3. The fermion free energy in the ideal gas limit.}\\[2mm]
\epsfig{file=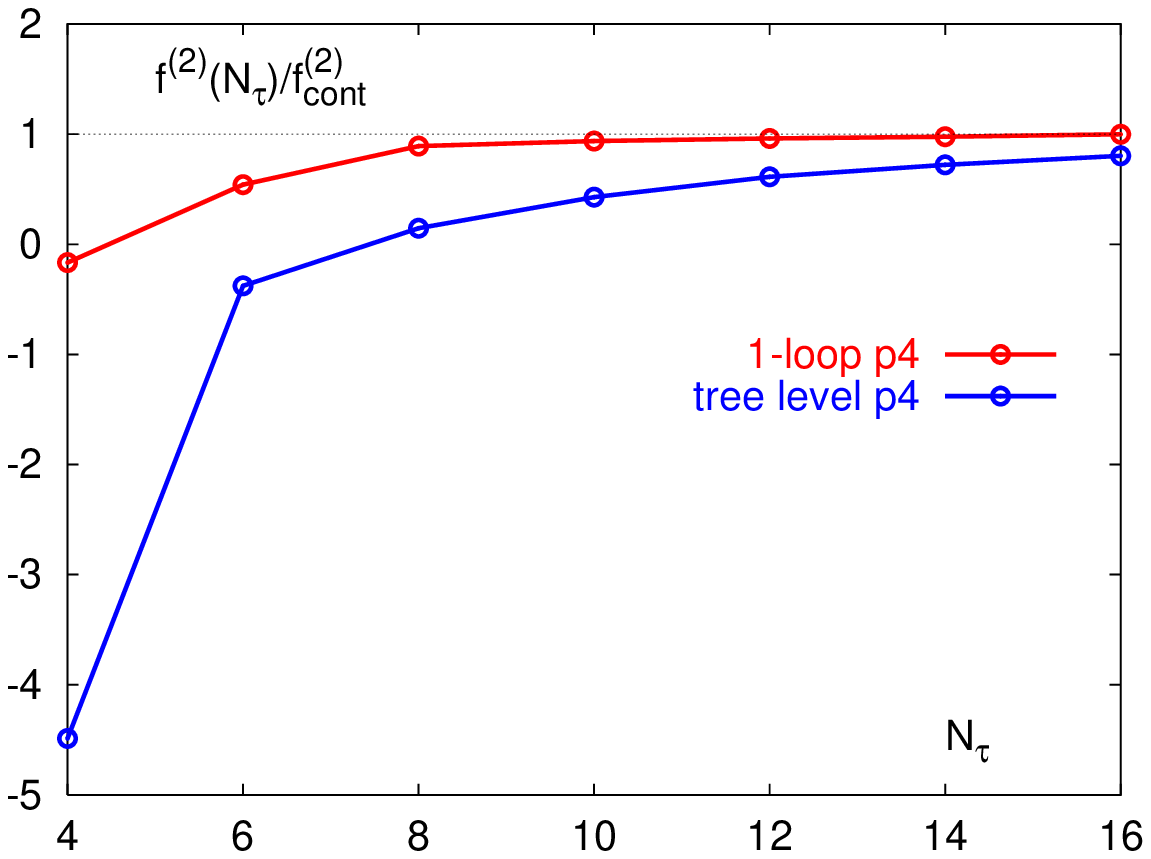,width=75mm}
{Figure 4. Corrections to the fermion free energy in one-loop order.}\\[1mm]
\section{THE SIMULATION}
We have performed simulations with the tree-level p4 action with a fat-weight $\omega =
0.2$ using the Hybrid R algorithm with a step size $\Delta \tau < m_qa/2$ and a
trajectory
  length of 0.8. We simulated three degenerate quark flavours of mass $m_qa = 0.1, 0.05$
  and $0.025$ on lattice sizes of $8^3 \times 4$ and $16^3 \times 4$.
\subsection{ROTATIONAL SYMMETRY}
To test the effect of improving the rotational symmetry of the action we
extracted the potential from Polyakov-Loop correlations according to
\bqas
V(R) &=& - {1\over N_{\tau}}~{\rm ln}~\left\langle PLC(R)\right\rangle\\
PLC(R) &=& \left\langle L^{\dag}(x)L(y)\right\rangle ,~~ {\rm with }
~R~=~|x-y|.
\eqas
The potential in figure 5 clearly shows string breaking for both actions. At
short distances the deviations from rotational symmetry are
smaller for the improved than for the standard action.\\[1mm]
\epsfig{file=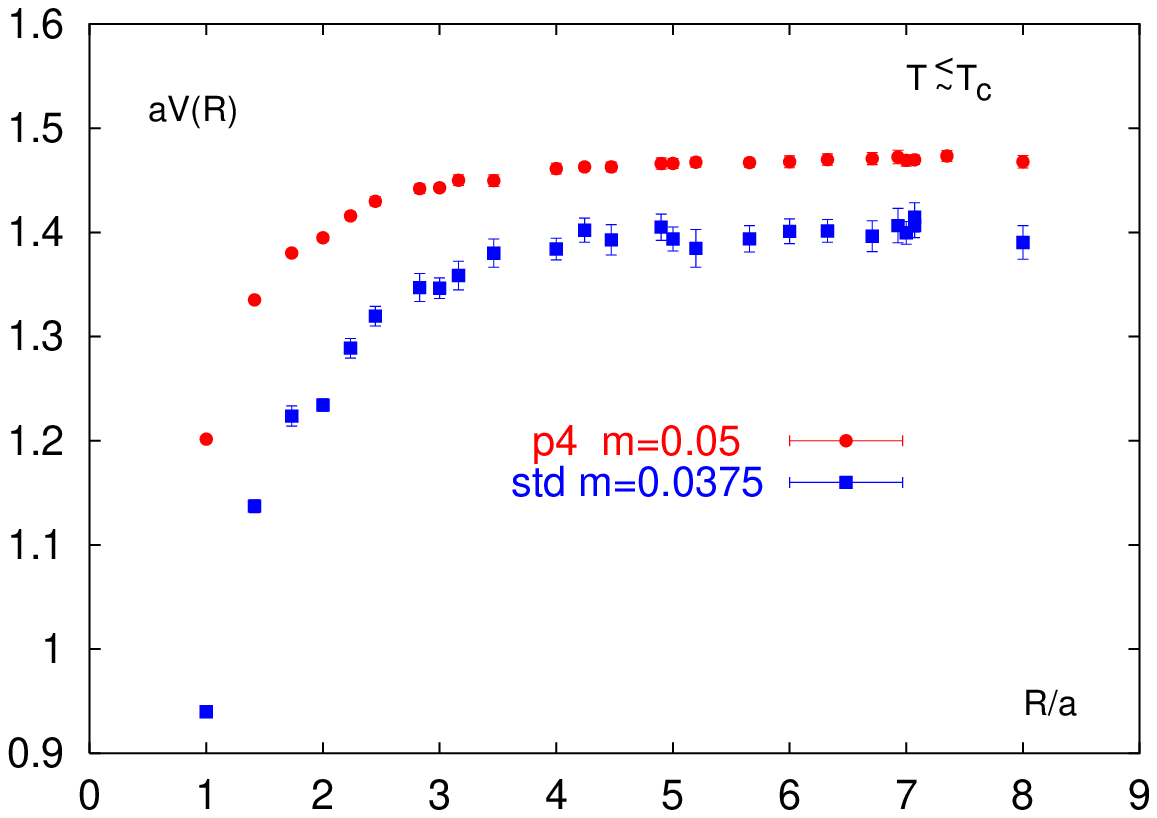,width=75mm}
{\vspace*{-4mm}{~\newline}}
{Figure 5. The potential from Polyakov loop correlations for the p4-
  and standard action\cite{deTar98}.}\\[2mm]
\subsection {CHIRAL PHASE TRANSITION}
In figure 6 we present our measurement of the chiral condensate. There is no sign
of a discontinuous behaviour in $\langle\bar{\Psi}\Psi\rangle$. Likewise there is no
two state signal in the time-histories. The finite-size effects are
small since \vspace*{2mm} the
\epsfig{file=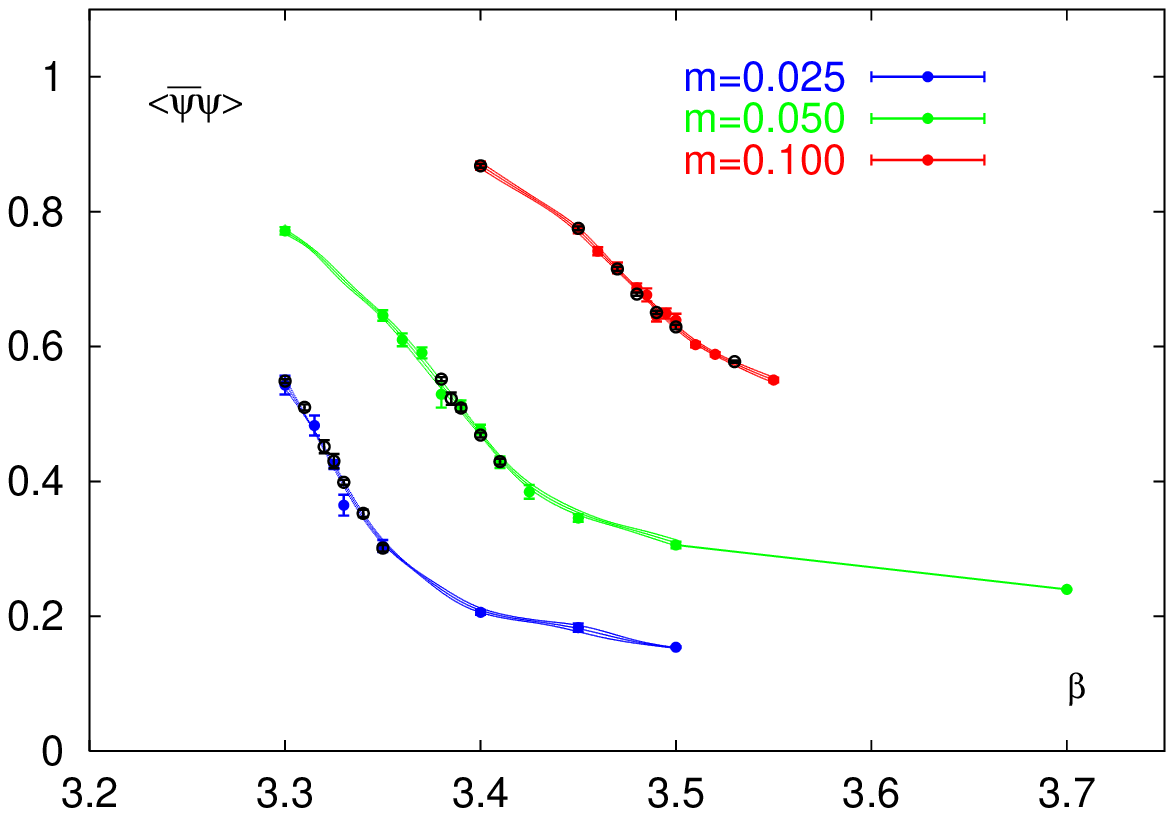,width=76mm}
{\vspace*{-2mm}{~\newline}}
{ Figure 6. The chiral condensate on $16^3 \times 4$ (black dots) and on $8^3
  \times 4$ (other symbols and curves). }\\
\epsfig{file=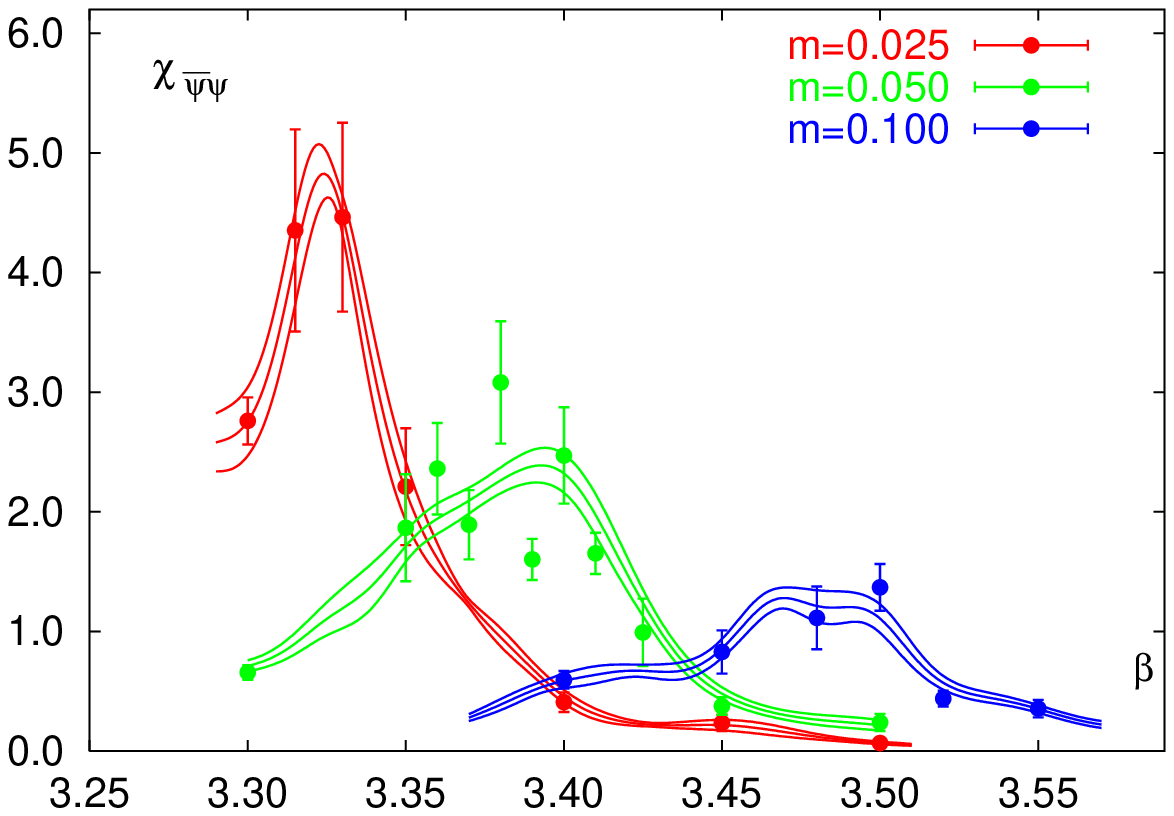,width=76mm}
{\vspace*{-1mm}{~\newline}}
{ Figure 7. The susceptibility of the chiral condensate on $8^3 \times 4$.}\\[2mm]
 results from the $16^3 \times 4$ lattice lie
perfectly on the  $8^3 \times 4$ curve. The increase of the peak height of the
susceptibility with smaller quark mass in figure 7 qualitatively agrees
with what has been found for the two flavour case. \\
We conclude that
there is no first order transition for bare quark masses of $m_q a=0.1$, $0.05$
and $0.025$. This confirms the earlier
findings with the standard staggered fermion action that a regime of first
order transitions only starts for quite light quarks.
On a qualitative level our calculation with an improved action gives the
impression that the first order regime is pushed to even smaller quark mass
values. This, however, requires a more detailed analysis of hadron masses in
order to set a physical scale.


\begin{thebibliography}{9}
\bibitem{wilczek92}
K. Rajagopal, F. Wilczek, Nucl. Phys. B399 (1993) 395 and references therein
\bibitem{brown90}
  F. Brown et al., Phys. Rev. Lett. 65 (1990) 2491
\bibitem{iwasaki96}
  Y. Iwasaki et al., Phys. Rev. D54 (1996) 7010
\bibitem{beinlich96}
  B. Beinlich et al., Phys. Lett. B390 (1997) 268
\bibitem{heller98}
  U. Heller, F. Karsch and B. Sturm, to be published
\bibitem{blum97}
  T. Blum et al., Phys. Rev. D55 (1997) 1133
\bibitem{peikert98}
  A. Peikert et al., Nucl. Phys. B (Proc. Suppl) 63 (1998) 895
\bibitem{deTar98}
  C. DeTar et al., HEP-LAT 9808028
\end{thebibliography}
\end{document}